\documentclass[traditabstract]{aa} 
\usepackage{graphicx}
\usepackage{natbib}
\usepackage{subfig}

\newcommand{\Mjup}{\ensuremath{\,{M}_{\rm Jup}}}                  
\newcommand{\Rjup}{\ensuremath{\,{R}_{\rm Jup}}}                  
\newcommand{\Teq}{\ensuremath{T_{\rm eq}}}                        
\newcommand{\safronov}{\ensuremath{\Theta}}                       
\newcommand{\kms}{\,km\,s$^{-1}$}                                 
\newcommand{\mss}{\,m\,s$^{-2}$}                                  
\newcommand{\pjup}{\ensuremath{\,\rho_{\rm Jup}}}                 

\usepackage{amsmath,amssymb,graphicx}

\begin{document}
%

\title{Kepler-539: a young extrasolar system with two
giant planets \\ on wide orbits and in gravitational interaction}
   \subtitle{}
\titlerunning{The planetary system Kepler-539}

   \author{
          L. Mancini \inst{1,2}
          \and
          J. Lillo-Box\inst{3,4}
          \and
          J. Southworth\inst{5}
          \and
          L. Borsato\inst{6}
          \and \\
          D. Gandolfi\inst{7,8}
          \and
          S. Ciceri\inst{1}
          \and
          D. Barrado\inst{3}
          \and
          R. Brahm\inst{9,10}
          \and
          Th. Henning\inst{1}}
    \institute{
    Max Planck Institute for Astronomy, K\"{o}nigstuhl 17, 69117 -- Heidelberg, Germany \\
    \email{mancini@mpia.de}
        \and
    INAF -- Osservatorio Astrofisico di Torino, via Osservatorio 20, 10025 -- Pino Torinese, Italy
        \and
    Depto. de Astrof\'{i}sica, Centro de Astrobiolog\`{i}a (CSIC-INTA), ESAC campus 28691 -- Villanueva de la Ca\~{n}ada, Spain
        \and
    European Southern Observatory, Alonso de Cordova 3107, Vitacura, Casilla 19001, Santiago de Chile, Chile
        \and
    Astrophysics Group, Keele University, Keele ST5 5BG, UK
         \and
    Dip. di Fisica e Astronomia ``Galileo Galilei'', Universit\`{a} di Padova, Vicolo dell'Osservatorio 2, 35122 -- Padova, Italy
         \and
    Dip. di Fisica, Universit\`{a} di Torino, via P. Giuria 1, 10125 -- Torino, Italy
         \and
    Landessternwarte K\"onigstuhl, Zentrum f\"ur Astronomie der Universit\"at Heidelberg, K\"onigstuhl 12, 69117 -- Heidelberg, Germany
         \and
    Instituto de Astrof\'{i}sica, Pontificia Universidad Cat\'{o}lica de Chile, Av. Vicu\~{n}a Mackenna 4860, 7820436 -- Macul, Santiago, Chile %
        \and
    Millennium Institute of Astrophysics, Av. Vicu\~{n}a Mackenna 4860, 7820436 -- Macul, Santiago, Chile
}
\abstract{We confirm the planetary nature of Kepler-539\,b (aka Kepler object of interest K00372.01), a giant transiting exoplanet orbiting a solar-analogue G2\,V star. The mass of Kepler-539\,b was accurately derived thanks to a series of precise radial velocity measurements obtained with the CAFE spectrograph mounted on the CAHA 2.2-m telescope. A simultaneous fit of the radial-velocity data and \emph{Kepler} photometry revealed that Kepler-539\,b is a dense Jupiter-like planet with a mass of $M_{\mathrm{p}} = 0.97 \pm 0.29 \, M_{\mathrm{Jup}}$ and a radius of $R_{\mathrm{p}} = 0.747 \pm 0.018 \, R_{\mathrm{Jup}}$, making a complete circular revolution around its parent star in 125.6 days. The semi-major axis of the orbit is roughly $0.5$\,au, implying that the planet is at $\approx 0.45$\,au from the habitable zone. By analysing the mid-transit times of the 12 transit events of Kepler-539\,b recorded by the \emph{Kepler} spacecraft, we found a clear modulated transit time variation (TTV), which is attributable to the presence of a planet c in a wider orbit. The few timings available do not allow us to precisely estimate the properties of Kepler-539\,c and our analysis suggests that it has a mass between $1.2$ and $3.6\,M_{\mathrm{Jup}}$, revolving on a very eccentric orbit ($0.4 < e \leq 0.6 $) with a period larger than 1000\,days. The high eccentricity of planet c is the probable cause of the TTV modulation of planet b. The analysis of the CAFE spectra revealed a relatively high photospheric lithium content, $A(\mathrm{Li})=2.48 \pm 0.12$\,dex, which, together with both a gyrochronological and isochronal analysis, suggests that the parent star is relatively young.
}



\maketitle

\section{Introduction}
\label{sec_1}

The transiting extrasolar planet (TEP) population turns out to be
the best \emph{atout} in the hands of exoplanetary scientists.
Thanks to the early ground-based systematic surveys (e.g.,
\citealp{bakos:2004,alonso:2004,mcCullough:2005,pollacco:2006,pepper:2007,alsubai:2013,bakos:2013})
and then to those from the space (CoRoT: \citealp{barge:2008},
\emph{Kepler}: \citealp{borucki:2011}), more than 1290 transiting
planetary systems have now been found. The possibility to easily
derive most of their physical and orbital parameters, and to
investigate even the composition of their atmosphere, make the
TEPs the most suitable targets to detect and study in detail. They
currently represent the best statistical sample to constrain the
theoretical models of planetary formation and evolution. Moreover,
transit time variation (TTV) studies allowed the detection of
additional non-transiting bodies in many TEP systems, highlighting
the great efficacy of high-precision and high-cadence photometry
in the search for extrasolar planets.

The majority of TEPs have been discovered by the \emph{Kepler} space telescope, which has revealed how varied they are, in terms of mass and size, and how diverse their architectures can be, confirming science-fiction pictures in several cases or going beyond human imagination in others. The numerous TEP discoveries achieved by {\it Kepler} are based on time-series photometric monitoring of the brightness of over 145\,000 main sequence stars, in which the periodic-dimming signal caused by a TEP can be found. However, such a signal can be mimic by other astrophysical objects in particular configurations. Radial velocity (RV) follow-up observations of the `potential' parent stars are therefore a fundamental step for distinguishing real TEPs from false positive cases.

Here we focus our attention on the \emph{Kepler} system Kepler-539.
Thanks to precise RV measurements, presented in Sect.\,\ref{sec_2}, we confirm the planetary nature of Kepler-539\,b (aka KOI-372\,b, K00372.01, KIC 6471021), a dense gas-giant planet moving on a wide orbit around a young and active G2\,V star ($V=12.6$\,mag). The analysis of the physical parameters of Kepler-539\,b and its parent star are described in Sect.\,\ref{sec_3}. Moreover, the clear variation observed in the mid-times transits of  Kepler-539\,b strongly support the existence of an additional massive planet, Kepler-539\,c, in the system, moving on a larger and very eccentric orbit, as discussed in Sect.\,\ref{sec_4}. We summarise our results in Sect.\,\ref{sec_5}.

\section{Observations and data analysis}
\label{sec_2}

\subsection{Kepler photometry}
\label{sec_2.1}

\begin{figure*}[th]
\begin{center}
\resizebox{\hsize}{!}{\includegraphics[angle=0]{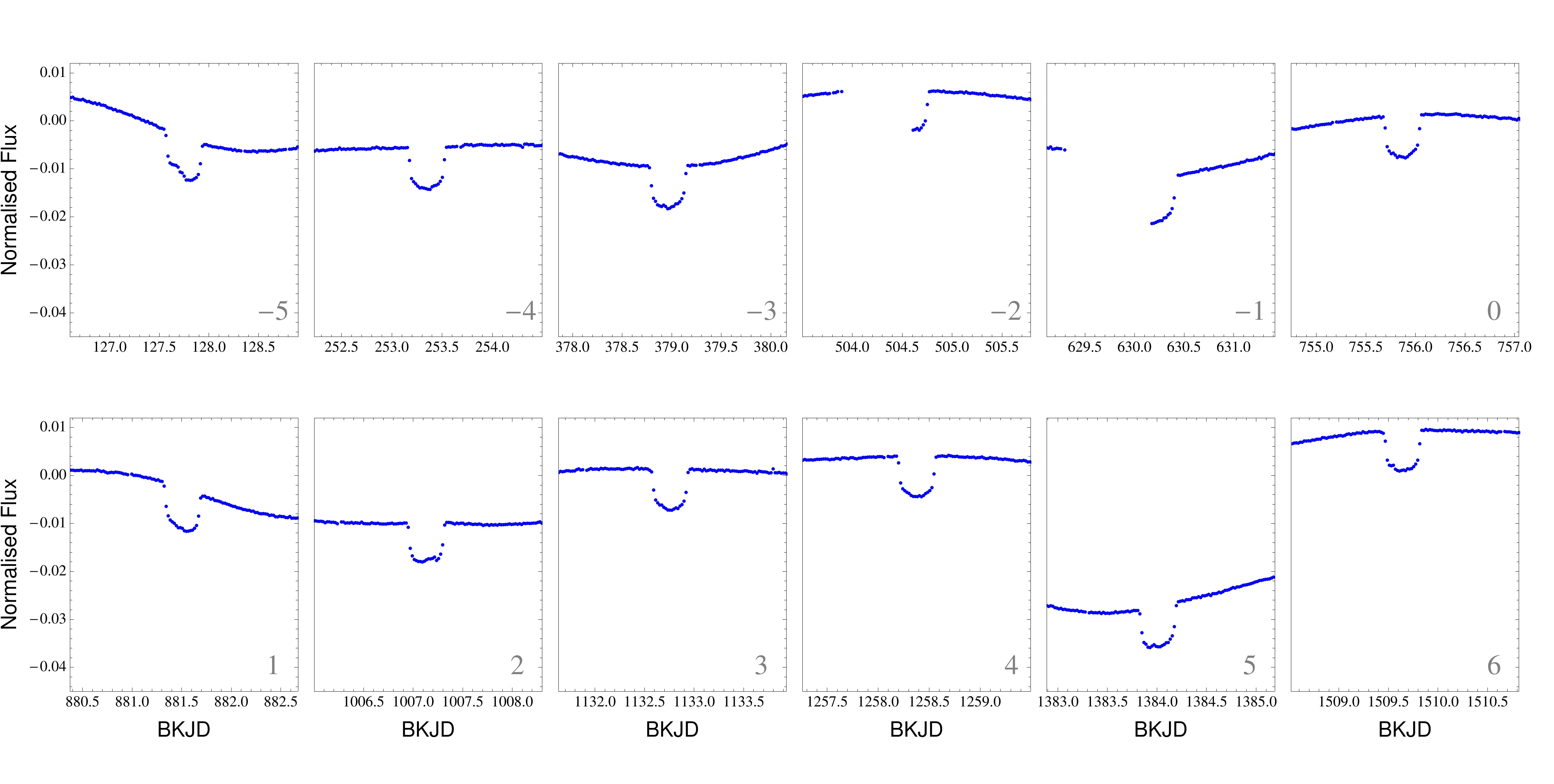}}
\caption{The twelve transit events of Kepler-539\,b observed by
\emph{Kepler} in long-cadence mode.  The transits at epoch -5 and 2 are clearly affected by star-spot-crossing events. Times are in BKJD (Barycentric Kepler Julian Date
-- equivalent to BJD(TDB) minus 2454833.0).}
\label{fig:Klc}
\end{center}
\end{figure*}

\begin{figure*}[th]
\begin{center}
\resizebox{\hsize}{!}{\includegraphics[angle=0]{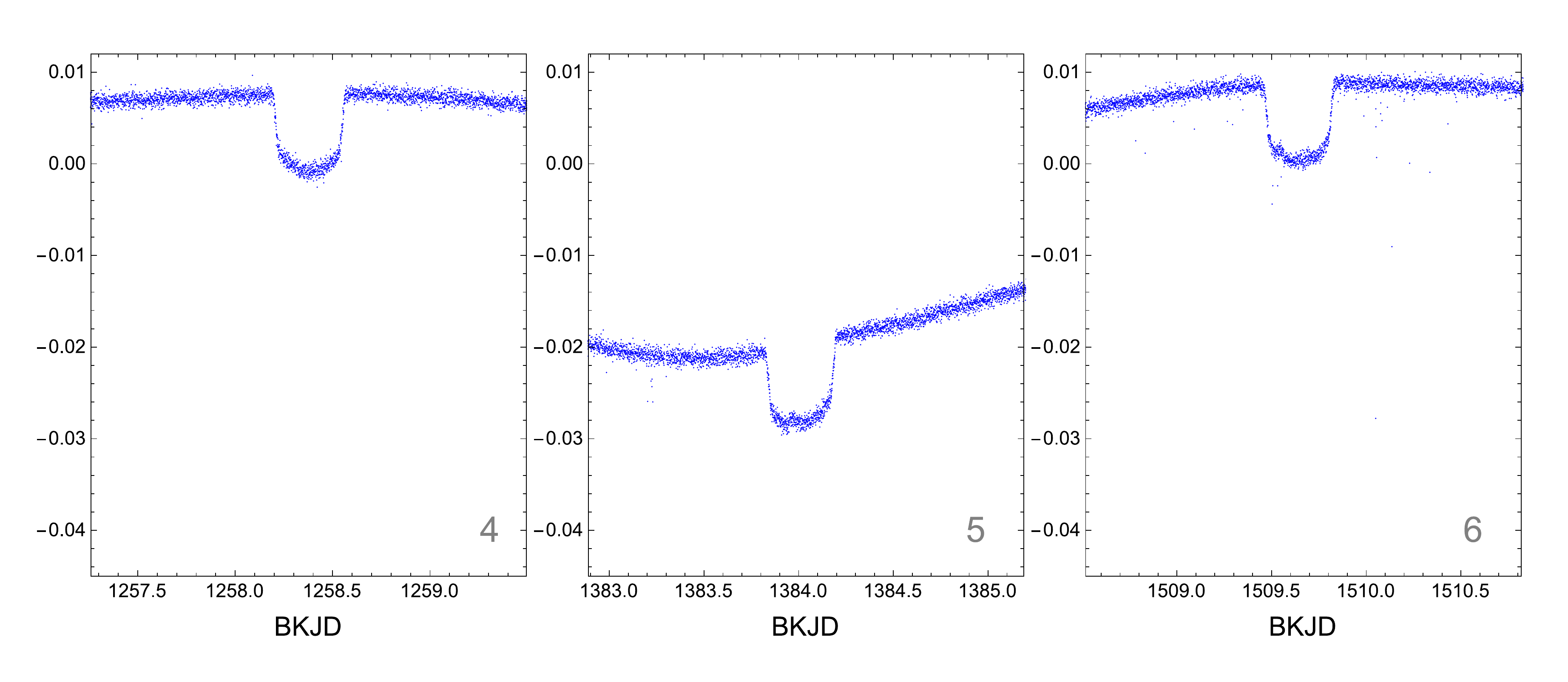}}
\caption{The three transit events of Kepler-539\,b observed by
\emph{Kepler} in short-cadence mode.  
Times are in BKJD (Barycentric Kepler Julian Date
-- equivalent to BJD(TDB) minus 2454833.0).}
\label{fig:Klc_sc}
\end{center}
\end{figure*}

\begin{figure*}[th]
\begin{center}
\resizebox{\hsize}{!}{\includegraphics[angle=0]{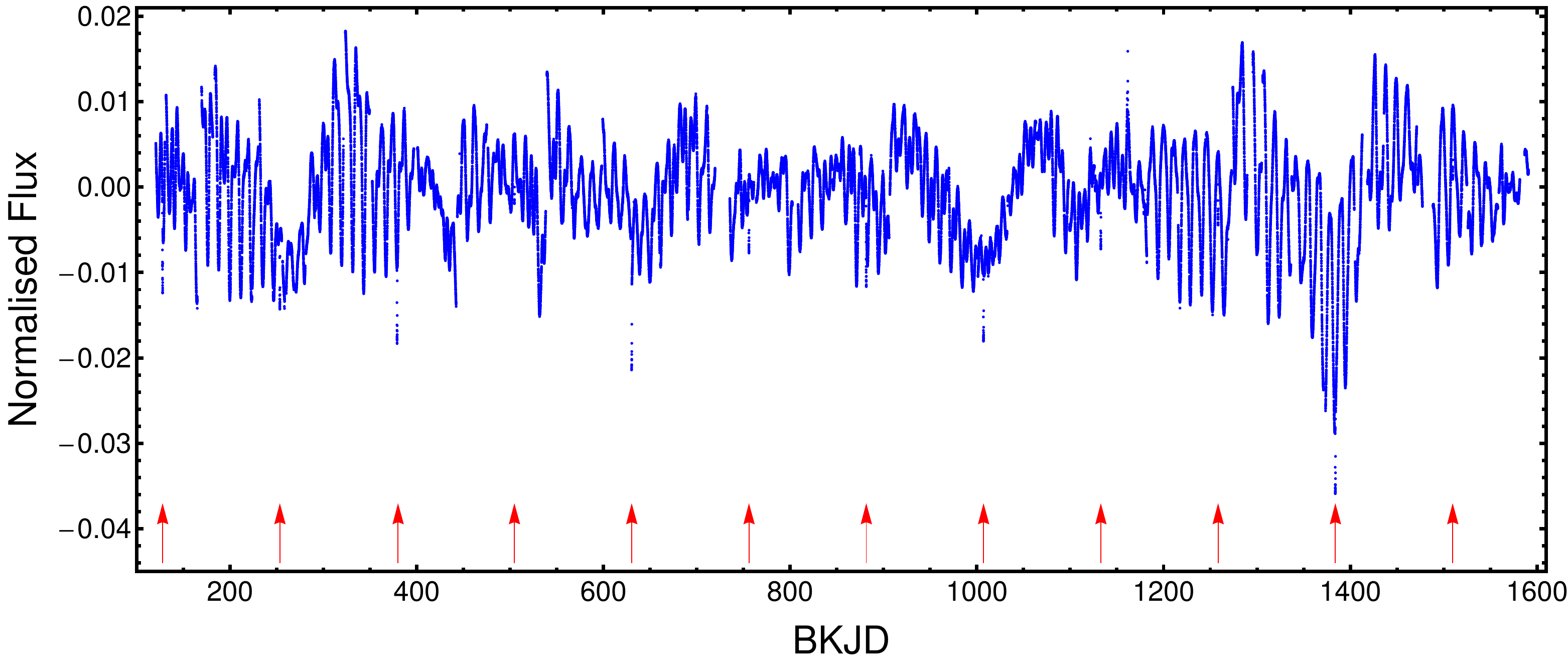}}
\caption{The entire \emph{Kepler} light-curve data of Kepler-539. The large stellar
variability can be reasonable interpreted as induced by a
star-spot activity. Times are in BKJD (Barycentric Kepler Julian Date
-- equivalent to BJD(TDB) minus 2454833.0). The red arrows mark the mid-times of the twelve transits of Kepler-539\,b.} \label{fig:KLCcomplete}
\end{center}
\end{figure*}

The \emph{Kepler} spacecraft monitored Kepler-539 from quarters 0 to
17 (i.e.\ four years; from May 2009 to May 2013). It was labelled
as a \emph{Kepler} object of interest (KOI) due to a $\sim 0.2\%$
dimming in its light curve with a period of $\sim 125$\,days
\citep{borucki:2011}. This periodic dimming is actually caused by
the transit of a Jupiter-like planet candidate, Kepler-539\,b, moving
on a quite wide orbit around the star. 12 transits of Kepler-539\,b
are present in the \emph{Kepler} long cadence (LC) light curves.
We have labelled them from cycle -5 to cycle 6 (see
Fig.\,\ref{fig:Klc}). Two of the transits are incomplete (cycles
-2 and -1); two are most likely contaminated by starspot crossing anomalies (cycles
-5 and 2); three were also covered in short cadence (SC) (cycles
4, 5 and 6; see Fig.\,\ref{fig:Klc_sc}). The complete \emph{Kepler} light curve is shown in
Fig.\,\ref{fig:KLCcomplete}, which highlights a significant
stellar variability ($0.0470 \pm 0.0002$\,mag peak-to-peak).
\citet{mcquillan:2013} found a periodic photometric modulation in
the light curve and, by assuming that it is induced by a star-spot
activity, estimated a stellar rotation period of $11.769 \pm
0.016$\,days. This value is in good agreement with that found by
\citet{walkowicz:2013a,walkowicz:2013b}, i.e.\ $11.90 \pm
3.45$\,days.


\subsection{Radial velocity follow-up observations}

We monitored Kepler-539 between July 2012 and July 2015 with the
Calar Alto Fibre-fed Echelle spectrograph (CAFE; \citealp{aceituno:2013})
mounted on the 2.2\,m telescope at the Calar Alto Observatory
(Almer\'ia, Spain) as part of our follow-up programme of {\it
Kepler} candidates. This programme has already confirmed the
planetary nature of Kepler-91\,b
\citep{lillo-box:2014a,lillo-box:2014b}, Kepler-432\,b
\citep{ciceri:2015}, Kepler-447\,b \citep{lillo-box:2015b}, and
has identified and characterised some false positives in the sample of \emph{Kepler} planet candidates
\citep{lillo-box:2015a}.
CAFE has a spectral coverage from 4000\,\AA\ to 9500\,\AA, divided into 84 orders with a mean resolving power of $R=63\,000$. We acquired 28 spectra, which were reduced using the dedicated pipeline provided by the observatory \citep{aceituno:2013}.
We used thorium-argon (ThAr) exposures obtained after each science spectrum to wavelength-calibrate the corresponding data. The final spectra have signal-to-noise ratios in the range S/N $=$ 7--16. A radial velocity (RV) was obtained from each spectrum by using the cross-correlation technique through a weighted binary mask \citep{baranne:1996}.
The mask is composed of more than 2000 sharp and isolated spectral
lines in the CAFE wavelength range. The cross-correlation was
performed in a $\pm 30$\,km\,s$^{-1}$ range around the expected RV
of the star. The peak of the cross-correlation function (CCF) was
measured by fitting a four-term Gaussian profile. This velocity
was then corrected for the barycentric Earth radial velocity at
mid-exposure time. Since we took several consecutive spectra on the same nights, we
decided to combine the RV values of the corresponding pairs in the
cases where their individual signal-to-noise was low (i.e., S/N
$<10$) and mutual discrepancies were larger than 50\,m\,s$^{-1}$.
This procedure can also diminish the effect of short term variability, such as the granulation noise, on the radial velocity (although here the expected
amplitudes are of the order of few tens of m\,s$^{-1}$). Some of the spectra were neglected due to low quality, mainly caused by weather conditions or problems with the stability of the ThAr lamp.
We obtained 20 RV values in total which are reported, together with their observing times, in Table~\ref{tab:RV} and are compatible with the presence of a $\sim 1\, M_{\mathrm{Jup}}$ planet in the system (see Fig.\,\ref{fig:rv} and Sect.\,\ref{sec_4}).

\begin{table}[h]
{\tiny
\centering %
\caption{RV and BVS measurements of Kepler-539 from CAHA/CAFE.}
\begin{tabular}{cccccc}
\hline %
\hline \\[-6pt] %
Date of observation    & RV             & err$_{\mathrm{RV}}$ & FWHM & BVS   \\
BJD(TDB)-2450000          & (km\,s$^{-1}$) & (km\,s$^{-1}$)   & (km\,s$^{-1}$)    & (km\,s$^{-1}$)     \\
\hline \\[-6pt] %
6116.5291378  &  10.408  & 0.026 & -0.041  & 9.696 \\ %
6124.3823828  &  10.383  & 0.029 & -0.065 & 9.517 \\ %
6138.4327769  &  10.387  & 0.018 & -0.117 & 9.656 \\ %
6523.4275256  &  10.402  & 0.024 & -0.100 & 9.572 \\ %
6598.2879800  &  10.448  & 0.033 & 0.081  & 9.518 \\ %
6804.5209537  &  10.439  & 0.029 & -0.146 & 9.250 \\ %
6811.4391381  &  10.475  & 0.035 & 0.159 & 9.407 \\ %
6821.6225006  &  10.432  & 0.049 & -0.118 & 9.198 \\ %
6834.4194255  &  10.437  & 0.025 & 0.086 & 9.342 \\ %
6859.5818459  &  10.409  & 0.038 & -0.391 & 10.036 \\ %
7136.6259467  &  10.285  & 0.039 & -0.266 & 9.541 \\ %
7152.5198831  &  10.431  & 0.042 & 0.087 & 9.422 \\ %
7160.5096750  &  10.450  & 0.021 & -0.267 & 9.312 \\ %
7169.5836757  &  10.500  & 0.032 & 0.038 & 9.153 \\ %
7193.4502198  &  10.425  & 0.045 & -0.088 & 9.556 \\ %
7254.4473639  &  10.361  & 0.035 & -0.021 & 9.727 \\ %
7258.3855966  &  10.400  & 0.045 & -0.106 & 9.559 \\ %
7263.4322174  &  10.402  & 0.028 & 0.014 & 9.519 \\ %
7264.4220946  &  10.413  & 0.025 & 0.018 & 9.599 \\ %
7266.3737337  &  10.449  & 0.034 & 0.169 & 9.995 \\ %
\hline %
\end{tabular}
\label{tab:RV}%
}
\end{table}

\subsection{Spectral analysis and the age of Kepler-539}
We derived the spectroscopic parameters of the host star Kepler-539
from the co-added CAFE spectra, which has a S/N ratio of about 40
per pixel at 5500\,\AA. Following the procedures described in
\citet{Gandolfi:2013,Gandolfi:2014}, we used a customised IDL\footnote{The acronym IDL
stands for Interactive Data Language and is a trademark of ITT
Visual Information Solutions.}
software suite to fit the composite CAFE spectrum to a grid of
synthetic theoretical spectra. The latter were calculated with the
stellar spectral synthesis program SPECTRUM \citep{Gray:1994}
using ATLAS9 plane-parallel model atmospheres \citep{kurucz:1979},
under the assumptions of local thermodynamic equilibrium (LTE) and
solar atomic abundances as given in \citet{Grevesse:1998}. We
fitted spectral features that are sensitive to different
photospheric parameters. Briefly, we used the wings of the Balmer
lines to estimate the effective temperature $T_\mathrm{eff}$ of
the star, and the Mg\,{\sc i} 5167, 5173, and 5184~\AA, the
Ca\,{\sc i} 6162 and 6439~\AA, and the Na\,{\sc i}~D lines to
determine the surface gravity $\log{g_{\star}}$. The iron
abundance [Fe/H] and microturbulent velocity $v_{\mathrm{micro}}$
was derived by applying the method described in
\citet{Blackwell:1979} on isolated Fe\,{\sc i} and Fe\,{\sc ii}
lines. To determine the macroturbulent velocity
$v_{\mathrm{macro}}$, we adopted the calibration equations for
solar like stars from \citet{doyle:2014}. The projected rotational
velocity $v\,\sin{i_{\star}}$ was measured by fitting the profile
of several clean and unblended metal lines\footnote{Here
$i_{\star}$ refers to the inclination of the stellar rotation axis
with respect to the line of sight.}. We found that Kepler-539 has an
effective temperature of $T_\mathrm{eff}=5820 \pm 80$\,K,
$\log{g_{\star}}=4.4 \pm 0.1$ (cgs), [Fe/H]$=-0.01\pm0.07$\,dex,
$v_ {\mathrm{micro}}=1.1\pm0.1$\,\kms,
$v_{\mathrm{macro}}=3.2\pm0.6$\,\kms, and
$v\,\sin{i_{\star}}=4.4\pm0.5$\,\kms
(Table~\ref{Parameter-Table}). According to the
\citet{Straizys:1981} calibration scale for dwarf stars, the
effective temperature of Kepler-539 translates to a G2\,V spectral
type.

\begin{figure}[th]
\begin{center}
\resizebox{\hsize}{!}{\includegraphics[angle=0]{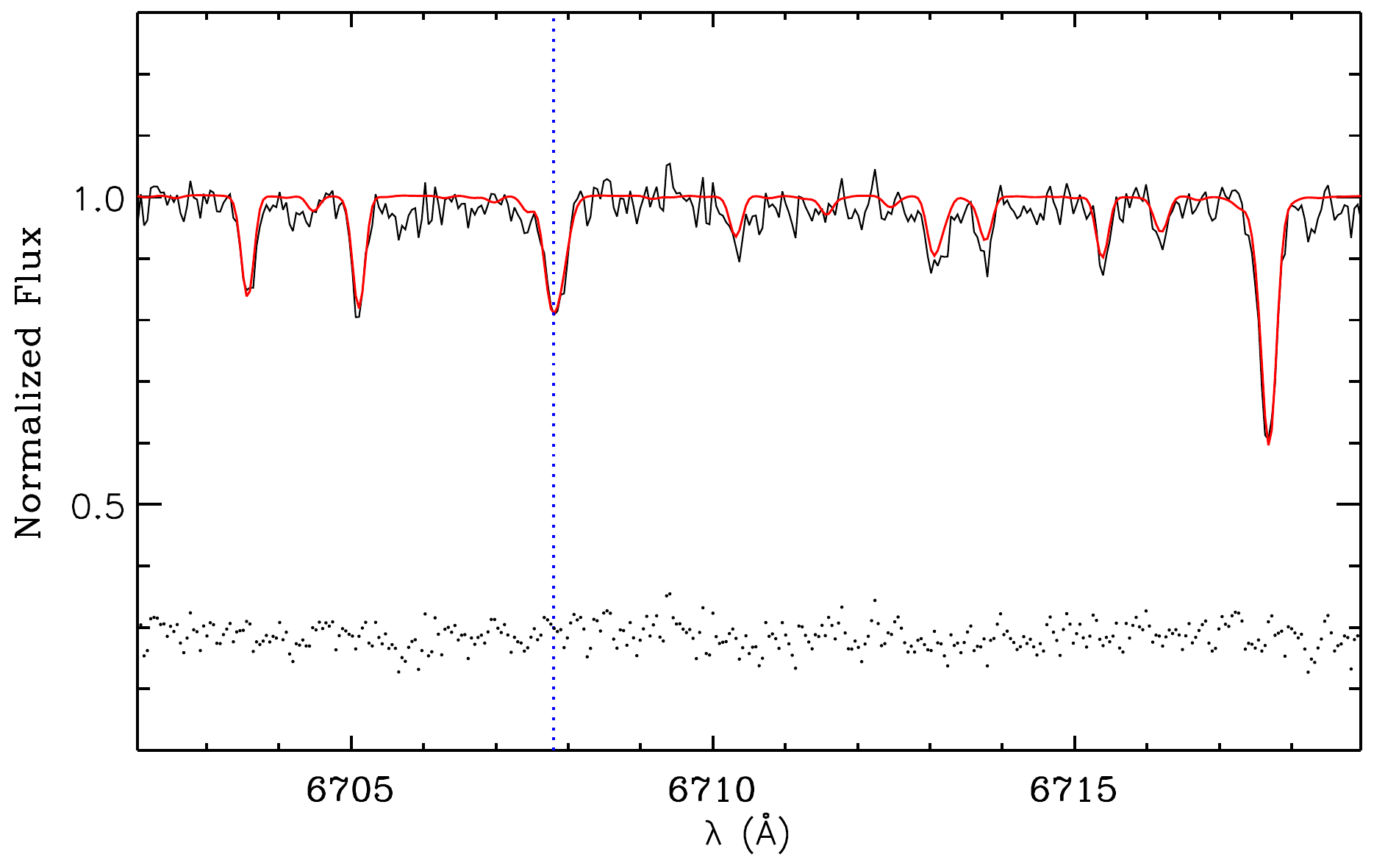}}
\caption{CAFE co-added spectrum of Kepler-539 (black line)
encompassing the Li\,{\sc i} 6707.8 \AA\ absorption doublet. The
best fitting ATLAS9 spectrum is overplotted with a thick red line.
The vertical dashed line marks the position of the Li doublet. The
lowest part of the plot displays the residuals to the fit.}
\label{fig:KOI372-Li}
\end{center}
\end{figure}

The CAFE co-added spectrum of Kepler-539 reveals the presence of a
moderate Li\,{\sc i} 6707.8 \AA\ absorption doublet
(Fig.~\ref{fig:KOI372-Li}). We estimated the photospheric lithium
abundance of the star by fitting the Li doublet using ATLAS9 LTE
model atmospheres. We fixed the stellar parameters to the values
given in Table~\ref{Parameter-Table} and allowed our code to fit
the lithium content. Adopting a correction for non-LTE effects of
+0.006\,dex \citep{Lind:2009}, we measured a lithium abundance of
$A({\rm Li})=\log\,(n({\rm Li})/n({\rm H}))+12=2.48\pm0.12$\,dex.

\begin{figure}[th]
\begin{center}
\resizebox{\hsize}{!}{\includegraphics[angle=0]{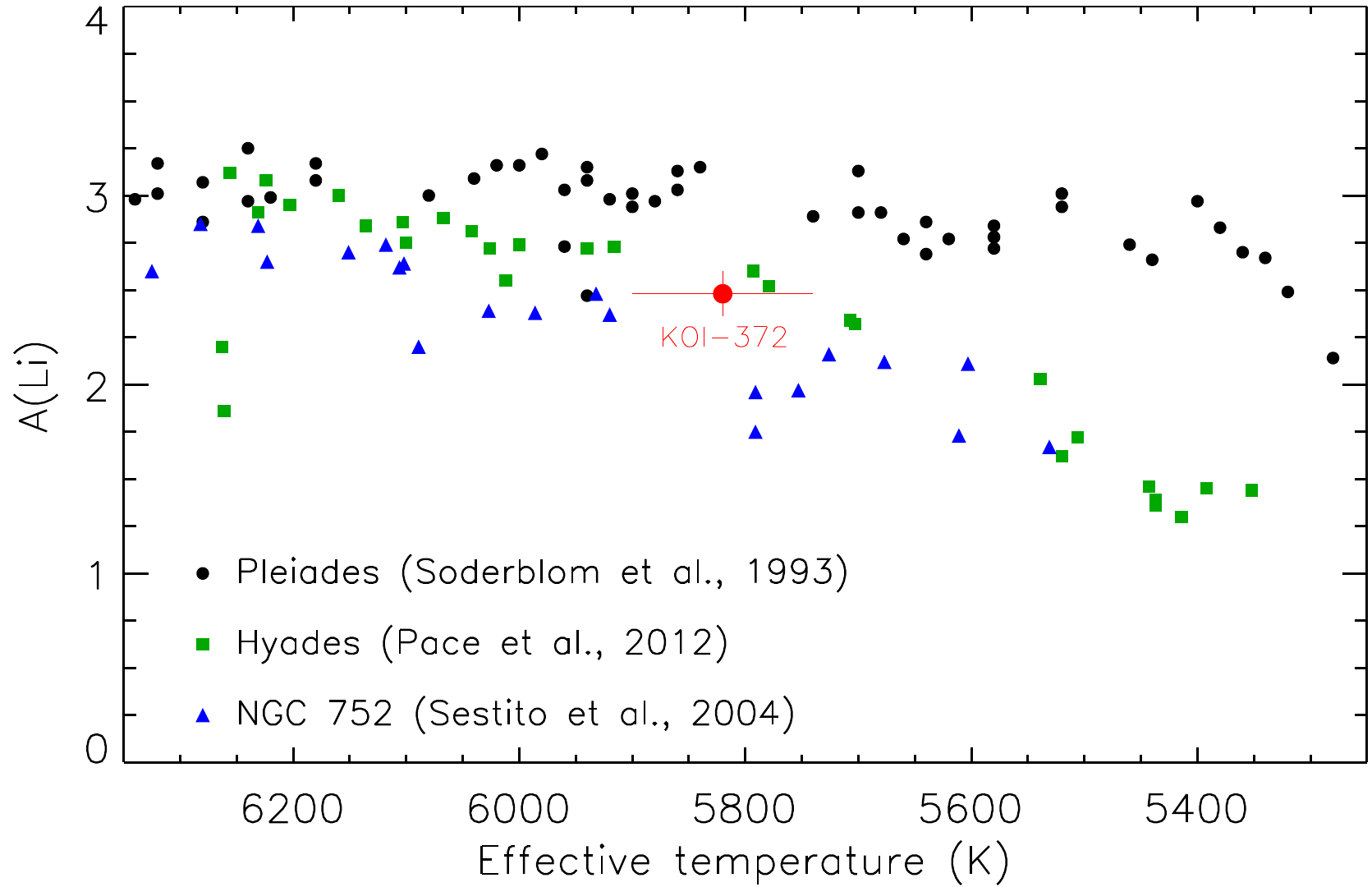}}
\caption{Lithium abundances of F- and G-type dwarf stars in the Pleiades
\citep[black dots;][]{Soderblom:1993}, Hyades
\citep[green squares;][]{Pace:2012}, and NGC\,752
\citep[blue triangles;][]{Sestito:2004} open clusters. The red dot
marks the position of Kepler-539.}
\label{KOI372-Li.vs.Teff}
\end{center}
\end{figure}

The photospheric lithium content and the moderate rotation
period of Kepler-539 suggest that the star is relatively young.
Fig.\,\ref{KOI372-Li.vs.Teff} shows the lithium abundance of Kepler-539
compared to the $A({\rm Li})$ of the dwarf stars of the
Pleiades, Hyades, and NGC\,752 open clusters, as listed in \citet{Soderblom:1993},
\citet{Pace:2012}, and \citet{Sestito:2004}, respectively. Kepler-539 has to be older
than $\sim$0.1~Gyr, as the star falls below the envelope of the Pleiades.
Although lithium depletion becomes ineffective beyond an age of 1--2 Gyr
\citep{Sestito:2005}, Kepler-539 lies between the envelopes of the other two clusters, suggesting
an age intermediate between the age of the Hyades ($\sim$0.6 Gyr) and the age of NGC\,752 ($\sim$2.0 Gyr).
This is further confirmed by the fact that the lithium content of Kepler-539
is intermediate between the average lithium abundance measured in early
G-type stars of 0.6-Gyr-old open clusters ($A({\rm Li})=2.58\pm0.15$\,dex) and
that of 2-Gyr-old open clusters \citep[$A({\rm Li})=2.33\pm0.17$\,dex;][]{Sestito:2005}.

We used Eq.~(32) from \citet{barnes:2010b} and the rotation period
of Kepler-539 to infer its gyrochronological age, assuming a convective turnover
time-scale of $\tau_\mathrm{c}=34$~days \citep{barnes:2010a} and a
zero-age main sequence rotation period of $P_0=1.1$\,days
\citep{barnes:2010b}. We found a gyrochronological age of $1.0\pm0.3$\,Gyr,
which supports the relatively young scenario. Our estimation is in
good agreement with the $1.15$\,Gyr gyrochronological age
predicted by \citet{walkowicz:2013a,walkowicz:2013b} and with that estimated using theoretical models (see Sect.\,\ref{sec_3} and Table\,\ref{Parameter-Table}).

By having short-cadence data over a quite large timespan, it is in principle possible to perform an asteroseismic analysis and try to precisely estimate the age of a star. However, in the case of Kepler-539, the SC data are only available for less than half the quarters. In particular, the continuous time baseline of SC data is not more than 125 d in the best case (BJD\,$2456015-2456139$). Furthermore the activity variation is predominant in the data. We analysed such SC data by filtering the variability due to activity with moving averages, but we did not find evidence of a clear frequency comb in the power spectrum and therefore we did not get particular clues about patterns due to solar-like oscillations.  We concluded that no reliable asteroseismic analysis could be performed for Kepler-539 based on the {\it Kepler} data.

\subsection{Excluding false-positive scenarios} %
A faint pulsating variable star or a binary system in the
background/foreground can mimic a planetary transit signal on the
target star. High-resolution images are extremely useful to
exclude this possibility (e.g., \citealp{lillo-box:2014c}). A
$K_{\mathrm{s}}$-band, high-resolution, adaptive optics image of
Kepler-539 was obtained with ARIES on the 6.5\,m MMT telescope by
\citet{adams:2012}, who found four stars within $6^{\prime\prime}$
of the target, see Table\,\ref{tab:dilution}. Following
\citet{lillo-box:2012}, we have estimated the dilution effect
caused by each companion, finding that it is very small for the
three closest targets, roughly 0.023\% in total and thus negligible. Concerning the
very bright companion ``E'', this is another object targeted as
KIC\,6471028. As it lies outside the \emph{Kepler} aperture, this
star does not contaminate Kepler-539. KIC\,6471028 was also detected
with the AstraLux North instrument mounted on the CAHA 2.2\,m
telescope by \citet{lillo-box:2012}, who estimated that this is a
K2-K4 background dwarf.

An intense stellar activity could mimic the presence of a
planetary body in the RV signal, thus causing a false positive
case. We have also analysed such a possibility by determining the
bisector velocity span (BVS) from the CAFE spectra (the values are
reported in Table\,\ref{tab:RV}). The bisector analysis provides a Pearson correlation coefficient between the RV and the BVS of $0.28^{+0.31}_{-0.42}$ (median value and $95\%$ confidence intervals from the posterior probability distribution, see \citealt{figueira:2016}). This value indicates only a weak correlation between the two parameters, which is only significant at a 2$\sigma$ level for the median value and still at less than 3$\sigma$ for the upper boundary. Consequently, we assume no correlation between both parameters, although we warn about the weak probability of correlation.

\begin{table}
\caption{Nearby visual companions around Kepler-539 (from
\citealp{adams:2012}) and
their dilution effect on the depth of the transit events.} %
\label{tab:dilution} %
\centering     %
\setlength{\tabcolsep}{8pt}
\begin{tabular}{ccccl}
\hline\hline\\[-6pt]
Companion & Distance\,$(^{\prime\prime})$ & $K_{\mathrm{s}}$\,(mag) & $\Delta K_{\mathrm{s}}$ & Dilution \\
\hline\\[-6pt]
B & 2.49 & 23.2 & 8.6 & 0.005 \% \\
C & 3.56 & 22.4 & 8.0 & 0.01~ \% \\
D & 4.99 & 22.7 & 8.2 & 0.008 \% \\
~E$^{\mathrm{a}}$ & 5.94 & 17.1 & 4.0 & 1.31~ \% \\
 \hline
\end{tabular}
\tablefoot{$^{\mathrm{a}}$Also known as KIC\,6471028.}
\end{table}
%

\section{Physical properties of the system}
\label{sec_3}
%
\begin{table*}
\begin{center}
\setlength{\tabcolsep}{4pt}
\caption{Final parameters of the planetary system KOI-0372.} %
\begin{tabular}{l c c c}
\hline
\hline  \\[-6pt]
Parameter   & ~~~~~ Nomen. ~~~~~ & ~~~~~~ Unit ~~~~~~ &  ~~~~~~~~~~~~~{\bf Value}$^{a}$ ~~~~~~~~~~~~~    \\
\hline \\[-6pt]%
\multicolumn{2}{l}{\emph{Stellar parameters}} \\
\noalign{\smallskip}
R.A. (J2000) \dotfill & \dotfill  & \dotfill & 19$^{\mathrm{h}}$56$^{\mathrm{m}}$29:39$^{\mathrm{s}}$ \\ %
Dec. (J2000) \dotfill & \dotfill  & \dotfill & 41$^{\circ}$52$^{\prime}$00.3$^{\prime \prime}$ \\ [3pt]%
Spectral type$^{b}$ \dotfill   &  \dotfill  & \dotfill & G2\,V              \\ [2pt]%
Kepler magnitude   \dotfill &  \dotfill $K_{\mathrm{p}}$  \dotfill &  \dotfill mag  \dotfill &$12.39$ \\ [3pt]%
Effective temperature      \dotfill & \dotfill $T_\mathrm{eff}$      \dotfill & \dotfill K    \dotfill & $5820  \pm 80$    \\
Iron abundance             \dotfill & \dotfill $[\mathrm{Fe/H}]$     \dotfill & \dotfill dex  \dotfill & $-0.01 \pm 0.07$  \\
Lithium abundance          \dotfill & \dotfill $A$(Li)               \dotfill & \dotfill dex  \dotfill & $2.5   \pm 0.1$    \\
Microturb.\ velocity       \dotfill & \dotfill $v_{\mathrm{micro}}$  \dotfill & \dotfill \kms \dotfill & $1.1   \pm 0.1$   \\
Macroturb.\ velocity$^{c}$ \dotfill & \dotfill $v_ {\mathrm{macro}}$ \dotfill & \dotfill \kms \dotfill & $3.2   \pm 0.6$   \\
Proj.\ rotat.\ velocity    \dotfill & \dotfill $v\,\sin{i_{\star}}$  \dotfill & \dotfill \kms \dotfill & $4.4   \pm 0.5$  \\
Stellar rotation period$^{d}$ \dotfill & \dotfill $P_\mathrm{rot}$ \dotfill & \dotfill day \dotfill   & $11.769 \pm 0.016$ \\ %
Stellar age (from gyrochronology) \dotfill & \dotfill  & \dotfill Gyr  \dotfill & $1.0   \pm 0.3$   \\ [2pt]%
Stellar age (from isochrones)   \dotfill & \dotfill  & \dotfill Gyr  \dotfill & $1.1_{-0.0\,-0.1}^{+0.6\,+0.3}$  \\
Stellar mass \dotfill                  & \dotfill $M_{\star}$ \dotfill & \dotfill $M_{\sun}$ \dotfill  & $1.048  \pm 0.034  \pm 0.025$   \\
Stellar radius \dotfill                & \dotfill $R_{\star}$ \dotfill & \dotfill $R_{\sun}$ \dotfill   & $0.952  \pm 0.017  \pm 0.007$   \\
Stellar mean density  \dotfill    & \dotfill $\rho_{\star}$ \dotfill & \dotfill $\rho_{\sun}$ \dotfill  & $1.216 \pm 0.054$    \\ %
Stellar surface gravity \dotfill   & \dotfill $\log{g_{\star}}$ \dotfill  & \dotfill cgs \dotfill & $4.502  \pm 0.014  \pm 0.003$  \\ [2pt]%
\hline \\[-6pt]%
\multicolumn{2}{l}{\emph{Planetary  parameters} (Kepler-539\,b)} \\
\noalign{\smallskip}
Planetary mass                    \dotfill & \dotfill $M_{\rm b}$    \dotfill & \dotfill \Mjup \dotfill & $0.97  \pm 0.29  \pm 0.02$  \\
Planetary radius                  \dotfill & \dotfill $R_{\rm b}$    \dotfill & \dotfill \Rjup \dotfill & $0.747 \pm 0.016 \pm 0.006$  \\
Planetary mean density            \dotfill & \dotfill $\rho_{\rm b}$ \dotfill & \dotfill \pjup \dotfill & $2.18  \pm 0.66  \pm 0.02$  \\ [2pt] %
Planetary surface gravity         \dotfill & \dotfill $g_{\rm b}$    \dotfill & \dotfill \mss  \dotfill & $43 \pm 12$             \\
Planetary equilibrium\ temperature  \dotfill & \dotfill \Teq           \dotfill & \dotfill K     \dotfill & $387.6 \pm   6.0$           \\
Safronov  number                  \dotfill & \dotfill \safronov\     \dotfill & \dotfill  & $1.24 \pm0.37 \pm 0.01$   \\ [2pt]%
\hline \\[-6pt]%
\multicolumn{2}{l}{\emph{Orbital parameters}} \\
\noalign{\smallskip}
Time of mid-transit      \dotfill & \dotfill $T_{0}$            \dotfill & \dotfill BJD$_{\mathrm{TDB}}$ \dotfill & $2\,455\,588.8710 \pm 0.0030$  \\ %
Orbital period           \dotfill & \dotfill $P_{\mathrm{orb}}$ \dotfill & \dotfill days                 \dotfill & $125.63243    \pm 0.00071$  \\ %
Semi-major axis          \dotfill & \dotfill $a$                \dotfill & \dotfill au  \dotfill & $0.4988 \pm 0.0054 \pm 0.0039$ \\ %
Orbital inclination              \dotfill & \dotfill $i$                \dotfill & \dotfill degree               \dotfill & $89.845   \pm 0.086$        \\ %
Fractional star radius   \dotfill & \dotfill $r_{\rm A}$        \dotfill & \dotfill                               & $0.01057  \pm 0.00078$     \\ %
Fractional planet radius \dotfill & \dotfill $r_{\rm b}$        \dotfill & \dotfill                               & $0.000854 \pm 0.000085$    \\ %
RV semi-amplitude        \dotfill & \dotfill $K_{\rm A}$        \dotfill & \dotfill m\,s$^{-1}$          \dotfill & $132.3    \pm 6.3$         \\ %
Barycentric RV           \dotfill & \dotfill $\gamma$           \dotfill & \dotfill km\,s$^{-1}$         \dotfill & $9.959    \pm 0.007$       \\ %
Eccentricity             \dotfill & \dotfill $e$                \dotfill & \dotfill & $<0.39^{e}$ \\ 
\hline
\end{tabular}
\tablefoot{Where there are two error bars, the first is a
statistical error, coming from the measured spectroscopic and
photometric parameters, while the second is a
systematic error and is given only for those parameters which have
a dependence on theoretical stellar models.
\\ $^{a}$ The adopted parameters assume a circular orbit.
\\ $^{b}$ With an accuracy of $\pm\,1$ sub-class.
\\ $^{c}$ Using the calibration equations of \citet{doyle:2014}.
\\ $^{d}$ From \citet{mcquillan:2013}.
\\ $^{e}$ The 99\% confidence upper limit on the eccentricity from a model in which
the eccentricity is allowed to vary in the fit.}
\label{Parameter-Table}
\end{center}
\end{table*}
\begin{figure}[th]
\resizebox{\hsize}{!}{\includegraphics[angle=0]{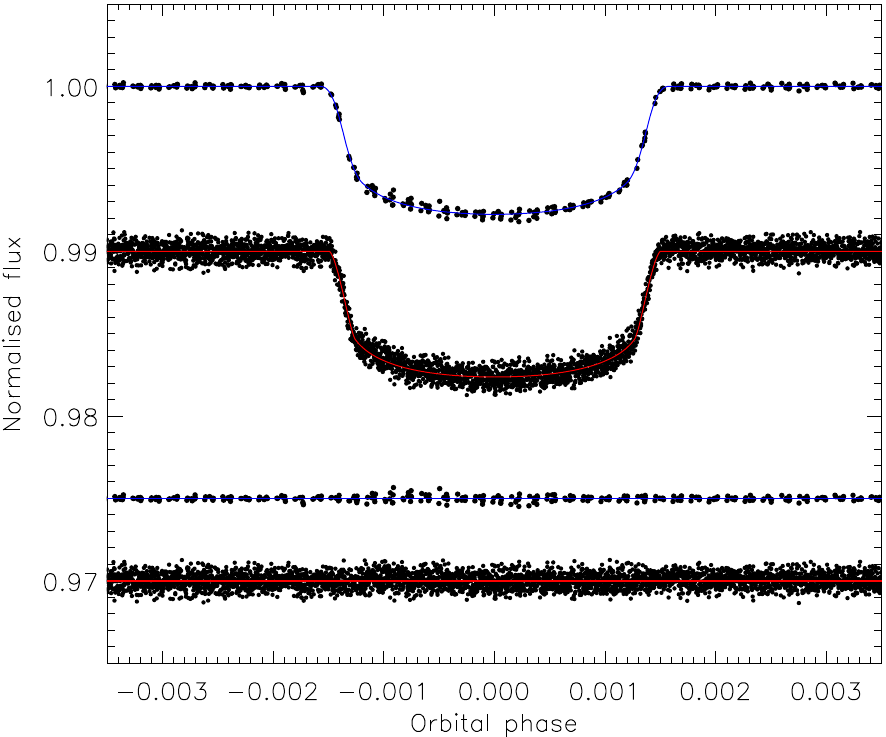}}
\caption{Phased Kepler long-cadence (top light curve) and
short-cadence (bottom light curve) data zoomed around transit
phase. The TTVs (see Sect.\,\ref{sec_4}) were removed from the data before plotting.
The {\sc jktebop} best fits are shown using solid lines. The
residuals of the fits are plotted at the base of the figure.}
\label{fig:lc}
\end{figure}
\begin{figure}[th]
\resizebox{\hsize}{!}{\includegraphics[angle=0]{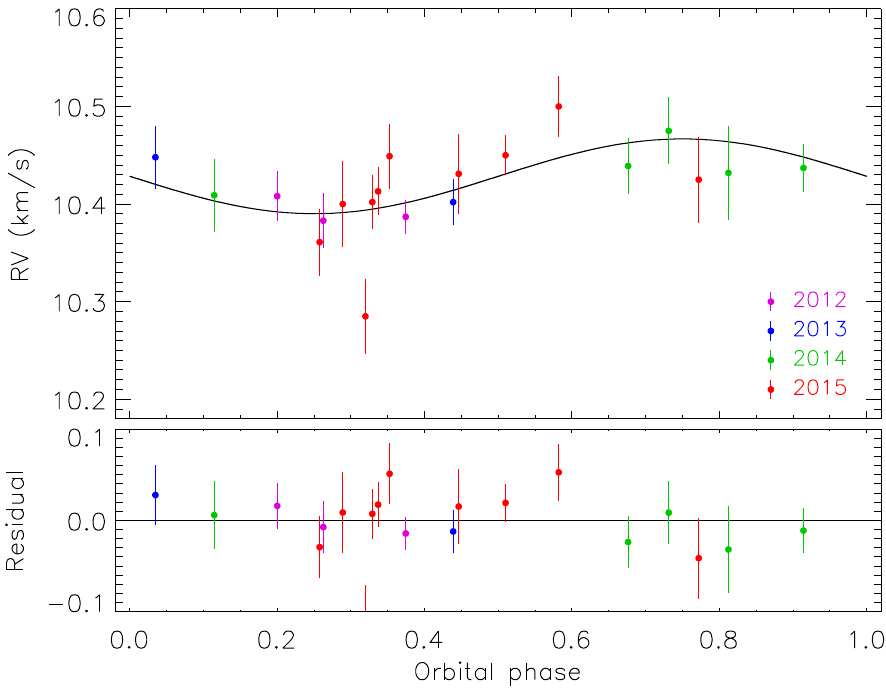}}
\caption{Upper panel: phased RVs for Kepler-539 and the best fit from
\textsc{jktebop}. Lower panel: residuals of RVs versus best fit.}
\label{fig:rv}
\end{figure}

For the determination of the physical parameters of the Kepler-539 system, we proceeded as in \citet{ciceri:2015}. We first selected for analysis all data within two transit durations of a transit, ignoring the two with only partial coverage in the {\it Kepler} light curve, and converted from flux to magnitude units. Each transit was then detrended by fitting a polynomial versus time. High polynomial orders of 3 to 5 were required to account for the brightness variations of the host star caused by spot activity.

We then fitted the photometry and the CAFE RV measurements simultaneously using the {\sc jktebop} code \citep{southworth:2013}, after modifying it to allow fitting for individual times of mid-transit. The {\it Kepler} LC and SC data were fitted separately, accounting for the long effective exposure times in the LC data by oversampling the fitted model by a factor of five.

The following parameters were fitted: the fractional radii of the two objects ($r_\star = R_\star/a$ and $r_{\rm p} = R_{\rm p}/a$, where $a$ is the orbital semi-major axis), the orbital inclination $i$, the time of midpoint of each transit, the velocity amplitude $K_\star$, the systemic velocity $V_{\gamma}$, and the coefficients of the polynomial for each transit. The orbital period, $P_{\mathrm{orb}}$, and reference transit midpoint, $T_0$, were fixed to the values measured from the transit times; these were used only for phasing the RV measurements. As in the other papers of our series (e.g. \citealt{mancini:2013a,mancini:2013b}), a quadratic limb darkening law was used, with the linear term fitted and the quadratic term fixed to 0.27 \citep{sing:2010}. We rescaled the error bars of the LC, SC and RV data to give a reduced $\chi^{2}$ of $\chi_{\nu}^{2} = 1.0$ for each versus the fitted model. Based on our experience with similar studies (e.g. \citealt{mancini:2014a,mancini:2014b}), this procedure is necessary for taking into account underestimated data errors and additional sources of white noise; for example an RV jitter due to the activity of the host star.

The best fits to the photometry and the RVs are shown in Figs.\ \ref{fig:lc} and \ref{fig:rv}. The uncertainties in the fitted parameters were derived by running both Monte Carlo and residual-permutation simulations \citep{southworth:2008} and choosing the larger of the two error bars for each parameter. The results obtained using the SC data are more precise than those from the LC data, so were adopted as the final set of photometric parameters.

We find that the observations are fully consistent with Kepler-539\,b moving on a circular orbit. However, we performed the fit both fixing the orbital eccentricity $e$ to zero and fitting for it via the combination terms $e\cos\omega$ and $e\sin\omega$, where $\omega$ is the argument of periastron.
In the latter case, we obtained $e=0.18^{+0.21}_{-0.18}$. We then used the Bayesian information criterion (BIC) to evaluate the preferred scenario, finding that the eccentric case is not favoured over the zero eccentricity model.

Finally, we estimated the age and the physical properties of the system from the SC results and the $T_{\rm eff}$ and [Fe/H] measured from our spectra. Constraints from theoretical stellar models were used to make the solution determinate. We followed the HSTEP approach \citep[][and references therein]{southworth:2012}, which yielded measurements of the properties of the star and planet accompanied by both statistical and systematic errorbars. The final results are reported in Table\,\ref{Parameter-Table}.

\section{Planet Kepler-539\,c from transit time variation}
\label{sec_4}

\begin{figure*}[th]
\resizebox{\hsize}{!}{\includegraphics[angle=0]{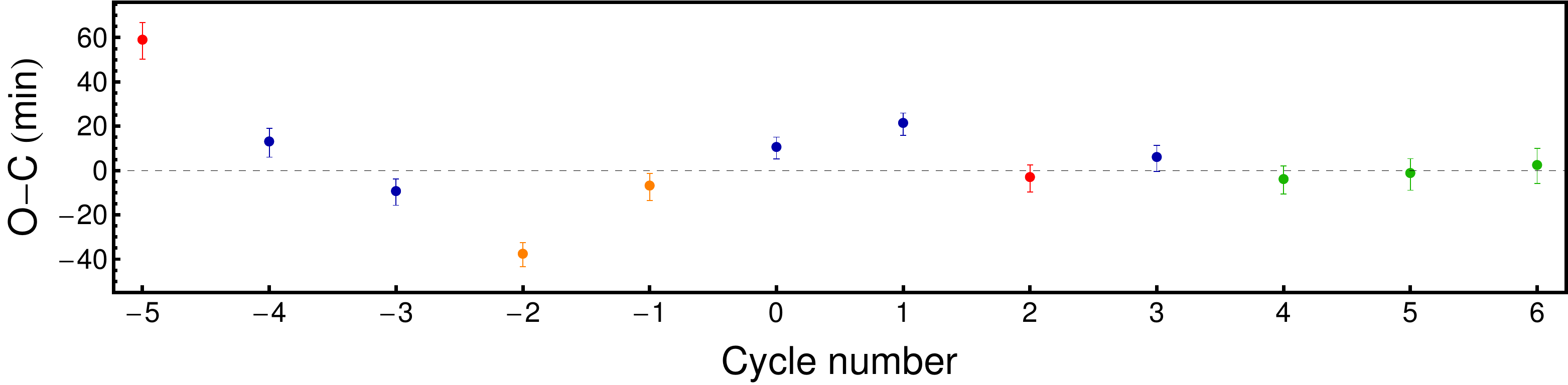}}
\caption{O--C diagram for the timings of Kepler-539\,b at
mid-transit versus a linear ephemeris. The timings in blue refer
to those coming from the \emph{Kepler} LC data, while those marked
with green are from SC data. The red points refer to the two LC
transits affected by star spots. The two incomplete LC transits
are marked with orange points.} \label{fig:OCplot}
\end{figure*}

We have analysed the 12 mid-transit times of Kepler-539\,b to characterise the ephemeris of the transit and check if there is a possible transit time variation (TTV), which could be a sign of the presence of additional bodies in this system. Since the transits at cycles -5 and 2 are quite affected by starspots, we used the {\sc prism}\footnote{Planetary Retrospective Integrated Star-spot Model.} and {\sc gemc}\footnote{Genetic Evolution Markov Chain.} codes \citep{tregloan:2013,tregloan:2015} that allowed us to fit both the full transit events and the shorter starspot-occultation events contemporaneously. The parameters of the starspots coming from the fits are reported in Appendix\,\ref{appendix_A}. The mid-transit times for the other epochs were estimated using {\sc jktebop}. The resulting timings are tabulated in Table\,\ref{tab:residuals} and were fitted with a straight line to obtain the orbital period, $P_{\mathrm{orb}}=125.63243 \pm 0.00071$\,days, and the reference time of mid-transit, $T_{0}=2\,455\,588.8710 \pm 0.0030$\,BJD\,(TDB). A plot of the residuals around the fit (see Fig.\,\ref{fig:OCplot}) shows a clear and modulated deviation from the predicted transit times, meaning that the orbital period of Kepler-539\,b is variable (the maximum TTV is $58.5 \pm 1.6$\,min and was measured for the transit at cycle $-5$). 
The uncertainties of $P_{\mathrm{orb}}$ and $T_{0}$ have been increased to account for this, but we stress that these values should be used with caution as they are based on only 12 timings, of which two are incomplete and others are affected by starspot anomalies that can introduce offsets in the transit timing measurements -- several studies have quantified this effect (e.g. \citealt{barros:2013,oshagh:2013,mazeh:2015,ioannidis:2016}). We have reported the predicted mid-transit times in Table\,\ref{tab:mid-transit-times} for the next year and a half.

\begin{table}
\caption{\emph{Kepler} times of transit midpoint of Kepler-539\,b and their residuals.} %
\label{tab:residuals} %
\centering %
\begin{tabular}{lcr}
\hline
\hline  \\[-6pt]
Time of minimum    & ~Cycle & Residual \\
BJD(TDB)$-2400000$ & ~no.   & (JD)~~~     \\
\hline \\[-6pt]%
$54960.74947  \pm   0.00109$ &    -5  &   0.040612  \\
$55086.35000  \pm   0.00040$ &    -4  &   0.008712  \\
$55212.96694  \pm   0.00042$ &    -3  &  -0.006781  \\
$55337.57977  \pm   0.00043$ &    -2  &  -0.026380  \\
$55463.23346  \pm   0.00118$ &    -1  &  -0.005126  \\
$55588.87808  \pm   0.00041$ &    ~0  &   0.007068  \\
$55714.51794  \pm   0.00043$ &    ~1  &   0.014495  \\
$55840.13340  \pm   0.00096$ &    ~2  &  -0.002478  \\
$55965.77212  \pm   0.00043$ &    ~3  &   0.003803  \\
$56091.39778  \pm   0.00030$ &    ~4  &  -0.002963  \\
$56217.00030  \pm   0.00030$ &    ~5  &  -0.001233  \\
$56342.66701  \pm   0.00030$ &    ~6  &   0.001403  \\
\hline
\end{tabular}
\end{table}
\begin{table}
\caption{The next mid-transit times of Kepler-539\,b.} %
\label{tab:mid-transit-times} %
\centering %
\begin{tabular}{ccllc}
\hline
\hline  \\[-6pt]
Cycle & BJD(TDB) &  ~~~~~Date & & Time (UT)    \\
\hline \\[-6pt]%
16 & 2457598.9899 & 2016 Jul & 29 &11:40 \\
17 & 2457724.6223 & 2016 Dec & 02 & 02:56 \\
18 & 2457850.2547 & 2017 Apr & 06 & 18:07 \\
19 & 2457975.8872 & 2017 Aug & 10 & 09:12 \\
20 & 2458101.5196 & 2017 Dec & 14 & 00:29 \\
\hline
\end{tabular}
\end{table}

In order to identify the origin of this modulated TTV, we used the TRADES (TRAnsits and Dynamics of Exoplanetary Systems; \citealp{borsato:2014}) code, which allows modelling of the dynamics of multiple-planet systems by reproducing the observed mid-transit times and RVs, with the possibility to choose among four different algorithms. 
We thus investigated the possibility that Kepler-539 is a planetary system formed by a star and two planets, b and c.
We ran two series of simulations, the first by selecting the PIKAIA algorithm \citep{charbonneau:1995}, the second with the Particle Swarm Optimization (PSO) algorithm \citep{tada:2007}. In all the cases, we fixed tight boundaries for the parameters of planet \textbf{b} (centred on the values in Table\,\ref{Parameter-Table}) and assumed the zero eccentricity case, which means that we fixed the argument of pericentre ($\omega_\textrm{b}$) to $90^{\circ}$,
and let only the mean anomaly ($\nu_\textrm{b}$) of planet b to freely float between $0$ and $360$ degrees.
We use a time of reference (or epoch, $t_\textrm{epoch} = 2455463.244$) that is very close to the transit cycle $0$, forcing the planet b to be initially at the pericentre. Because of this, all the solutions of both the algorithms returned zero values for $\nu_\textrm{b}$. Furthermore, we assumed for simplicity the two planets to be coplanar. We investigated different configurations of two planet system, with the period of planet c, $P_\textrm{c}$,  varying from $150$ to $2500$ days. We fitted the pair $(e_\textrm{c} \cos \omega_\textrm{c}, e_\textrm{c} \sin \omega_\textrm{c})$ instead of $(e_\textrm{c}, \omega_\textrm{c})$ to avoid correlation between the parameters. We set $0.9$ as the highest value allowed for the eccentricity,  while we let $\omega_\textrm{c}$ and $\nu_\textrm{c}$ freely vary as for $\nu_\textrm{b}$. The solutions obtained from the TRADES fitting processes were further analysed using the Frequency Map Analysis (FMA;  \citealp{laskar:1992}) to check if they are stable.

We found that the best solutions are those related to configurations with $P_\textrm{c} > 1000$\,days and a mass of the planet c between $1.2$ and $3.6\,M_\textrm{Jup}$ on a very eccentric orbit ($0.4 < e_\textrm{c} \le 0.6)$. The final parameters, with confidence intervals estimated with a bootstrapping method, are reported in Table\,\ref{tab:trades} for the five best solutions, based on the $\chi^2$, and displayed in figures shown in Appendix\,\ref{appendix_B}. The estimated confidence intervals are very tight to the fitted values because the algorithms found solutions that correspond to minimum surrounded by very high peaks in the $\chi^{2}$-space. We also exhaustively investigated narrow region ($200-800$\,days) of $P_\textrm{c}$ (in steps of about $300$ days), but the solutions that we obtained are not favoured because the corresponding $\chi^2$ is much higher than those reported in Table\,\ref{tab:trades}.

Knowing the orbital period of Kepler-539\,c, we can easily estimate the semi-major axis and the equilibrium temperature of the planet c for each solutions. These values are shown in right-hand columns of Table\,\ref{tab:trades}. Moreover, we can use Eq. (7.29) from \citet{haswell:2010} for estimating the maximum TTV signal expected for Kepler-539\,b, i.e.
\begin{equation}
\mathrm{TTV}_{\rm max} = (M_{\rm A}+M_{\rm b}) e_{\mathrm c} (a_{\mathrm b}/a_{\mathrm c})^{3}P_{\mathrm c}.
\end{equation}
The last column of Table\,\ref{tab:trades} reports the values of TTV$_{\rm max}$ for each of the five solutions. The best TRADES solution is also that for which the maximum TTV signal is higher than the others and in better agreement with the observations.

Very interestingly, the high eccentricity of planet c gives a direct explanation of the modulation of the TTV of planet b: when planet c moves close to planet b, it gravitationally kicks its smaller sibling, causing the maximum in the TTV signal; after the conjunction, planet b re-circularizes own orbit very slowly, because of tidal interactions with the star and planet c moving far from it. This causes the decreasing of amplitude of its TTV that we see in Fig\,\ref{fig:OCplot}.

We again stress that all the values presented in Table\,\ref{tab:trades} should be taken with caution, because the existence of planet c can be definitively constrained only by new transit observations, able to completely cover the TTV phase of planet b. In particular, one should verify if there is a periodic repetition of the peak and the modulation.
\begin{table*}
{\tiny
\caption{{\it Left-hand columns}: parameters of Kepler-539\,c from the best fit of the \emph{Kepler} mid-transit times and CAFE-RV measurements. The best five solutions obtained with TRADES are shown. {\it Left-hand columns}: parameters determined from the previous.}
\label{tab:trades} %
\centering %
\begin{tabular}{l c c c c c c  | c c c}
\hline %
\hline  \\[-6pt]
Algortihm  & $M_{\rm c}$\,(\Mjup) & $P_{\mathrm{c}}$\,(day) & $\nu_{\rm c}$\,($^{\circ}$) & $e_{\rm c}$ & $\omega_{\rm c}$\,($^{\circ}$) & $\chi_{\nu}^2$ & $a$\,(au) & $T_{\rm eq} \,(K)$ & TTV$_{\rm max}$\,(min) \\
\hline \\[-6pt]%
PIKAIA & $3.60_{-0.23}^{+0.30}$ & $1040_{-19}^{+30}$ & $~33.95_{-1.02}^{+0.76}$ & $0.432_{-0.006}^{+0.010}$ & $270.67_{-0.06}^{+0.05}$ & 34.2 & $2.04 \pm 0.04$ & $271 \pm 4$ &$31.0 \pm 3.6$ \\ [2pt]%
PSO & $1.27_{-0.05}^{+0.05}$ & $1705_{-49}^{+70}$ & $~23.13_{-0.72}^{+0.84}$ & $0.605_{-0.005}^{+0.009}$ & $166.27_{-0.86}^{+0.83}$ &39.5 & $2.84 \pm 0.08$ & $230 \pm 4$ & $9.3 \pm 1.1$\\ [2pt]%
PIKAIA & $1.20_{-0.11}^{+0.11}$ & $945_{-6}^{+7}$ & $~18.00_{-0.60}^{+0.57}$ &$0.448_{-0.010}^{+0.010}$ & $329.27_{-2.54}^{+2.69}$ & 57.2 & $1.91 \pm 0.02$ & $280 \pm 3$& $11.8 \pm 1.4$\\ [2pt]%
PIKAIA & $2.72_{-0.19}^{+0.15}$ & $1784_{-59}^{+66}$ & $349.14_{-0.57}^{+0.50}$ & $0.563_{-0.010}^{+0.010}$ & $296.56_{-1.70}^{+2.17}$ & 62.9 & $2.92 \pm 0.08$ & $226 \pm 4$& $17.8 \pm 2.2$\\ [2pt]%
PSO & $2.28_{-0.16}^{+0.15}$ & $963_{-27}^{+29}$ & $355.50_{-0.38}^{+0.34}$ & $0.457_{-0.009}^{+0.009}$ & $343.54_{-1.05}^{+1.34}$ & 66.6 & $1.94 \pm 0.05$ & $278 \pm 4$ & $22.4 \pm 2.7$\\
 \hline
\end{tabular}
\tablefoot{We used two different algorithm: the PIKAIA
algorithm \citep{charbonneau:1995} and the Particle Swarm
Optimization (PSO) algorithm \citep{tada:2007}, respectively. The first five columns refer to the values of the parameters determined from the fitting processes, while the others values have been determined using the previous ones. In particular, $\delta t$ represents the maximum TTV for the inner planet b.}
}
\end{table*}

\section{Discussion and conclusions}
\label{sec_5}

Thanks to precise RV measurements obtained with the high-resolution spectrograph CAFE, we confirmed the planetary nature of Kepler-539\,b, a dense Jupiter-like planet with a mass of $0.97 \pm 0.29\,\Mjup$ and a radius of $0.747 \pm 0.018\,\Rjup$, revolving with a period of $125.6$\,days on a circular orbit around a G2\,V star, similar to the Sun. The parameters of the parent star and the planet b were obtained by analysing the CAFE spectra and through a joint fit to the RV data and \emph{Kepler} transit light curves. Orbiting at $\sim 0.5$\,au from its host, Kepler-539\,b is located not so far ($\approx 0.45$\,au) from its \emph{habitable zone} (HZ), which has a width of $\sim 0.72$\,au, as we estimated it based on the stellar parameters reported in Table\,\ref{Parameter-Table} and the HZ calculator by \citet{kopparapu:2013,kopparapu:2014}. The parent star is quite active, as is shown by the $0.047$\,mag peak-to-peak modulation present in the long time-series photometry and from the star-spot anomalies clearly visible in two of the transit events monitored by \emph{Kepler}. Even though it has physical characteristics that resemble those of the Sun, Kepler-539 is much younger: $1.0 \pm 0.3$\,Gyr as estimated from its lithium abundance and gyrochronology.
\begin{figure*}[th]
\resizebox{\hsize}{!}{\includegraphics[angle=0]{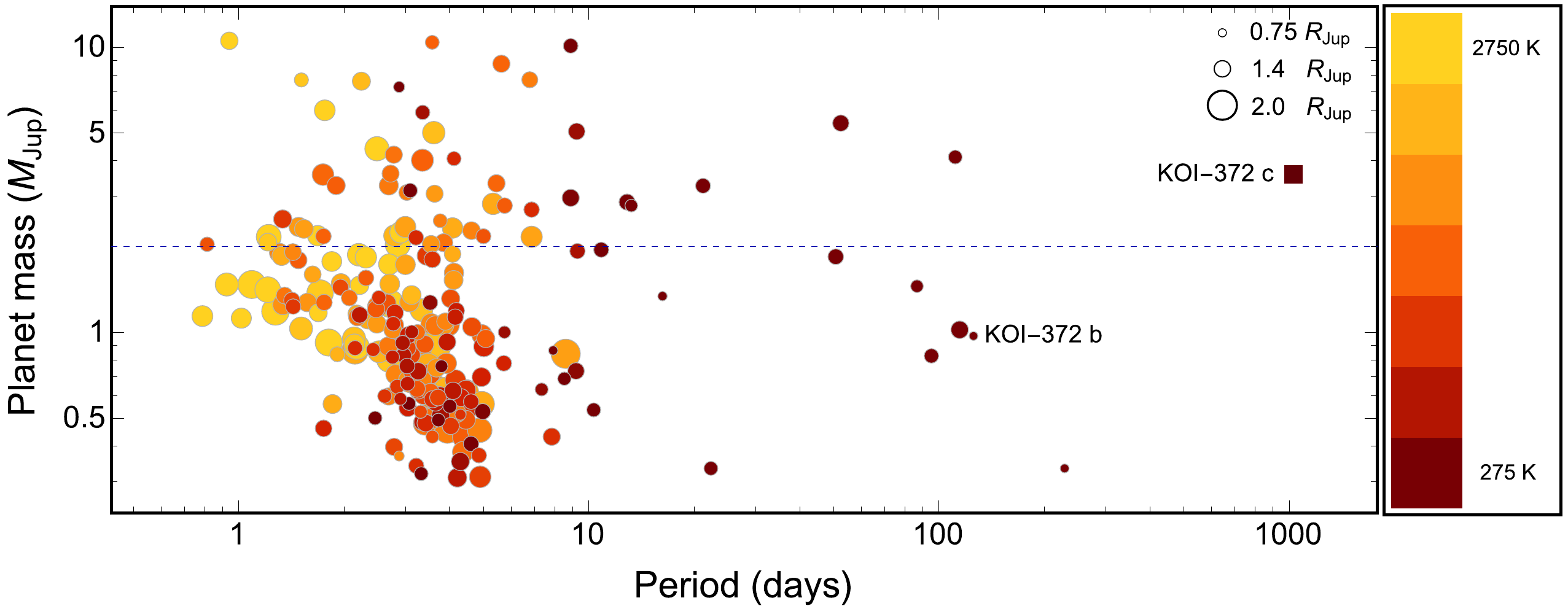}}
\caption{The \emph{period--planetary mass} diagram for
transiting planets in the mass range $0.3\,M_{\rm Jup}<M_{\rm p}<11\,M_{\rm Jup}$. The size of each circle is proportional to the
corresponding planetary radius, while the colour indicates equilibrium temperature (data taken from TEPCat). The dashed line demarcates the high-mass regime ($M_{\rm p} > 2 M_{\rm Jup}$). The positions of Kepler-539\,b and Kepler-539\,c are highlighted. Kepler-539\,c is marked with a box since we do not know its radius. The error bars have been suppressed for clarity.}
\label{fig:diagram}
\end{figure*}

An analysis of the \emph{Kepler} mid-transit times of Kepler-539\,b revealed a clear modulated TTV signal. We studied this with TRADES, a code able to make a simultaneous fit of RV and mid-transit times and compare the results with simulated data of various multi-planetary configurations. We found that the TTV can be explained through the presence of a $1.2$--$3.6\,M_{\mathrm{Jup}}$ Jovian-like planet c on a very eccentric ($e=0.43$--$0.61$) and wider orbit than that of Kepler-539\,b, with a period larger than 1000 days. Each of the five solutions that we have presented are related to a planetary system which is stable according to the results of an FMA analysis.

Since the orbit of Kepler-539\,c is highly eccentric, then the modulation of the TTV signal of Kepler-539\,b can be easily explained. As the distance of planet c from the parent star varies with time, there is a periodic change of the mid-transit time of planet b. In particular, when the planet c is in conjunction with planet b, the gravitational interaction between the two planets is maximum and we see a large amplitude of the TTV signal. Then, when planet c moves away, planet b re-circularizes its orbit very slowly and we see the gradual decreasing of the amplitude of its TTV.

Given its relatively young age, the physical parameters of the Kepler-539 planetary system can be very precious for astrophysicists working on theories of planet formation and evolution. If we visualise the known transiting planets, reported in TEPCat\footnote{TEPcat (Transiting Extrasolar Planet Catalogue) is available at www.astro.keele.ac.uk/jkt/tepcat/ \citep{southworth:2011}.}, in a period--mass diagram, we can see that Kepler-539\,b and Kepler-539\,c occupy sparsely-populated regions of this plot (here for the mass of Kepler-539\,c we have considered the best value found by TRADES).

We also underline that the transits of Kepler-539\,b are quite deep (0.8\%) and that the host star is relatively bright ($V=12.5$\,mag) and therefore amenable for exoplanet atmosphere studies. The star-planet distance is large enough to consider Kepler-539\,b as a non-inflated planet with a low irradiation level. Since there are only a handful of transiting Jupiter-like planets with a low irradiation level, this planetary system could be a precious target for future studies of exoplanet atmospheres of normal giant gas planets.

Unfortunately, the parameters of Kepler-539\,c are not well constrained as they are based on a poorly-sampled TTV (only 12 transit timings). More transits of Kepler-539\,b and more precise RV observations with a larger-aperture telescope are required to properly model its TTV signal and accurately determine the existence and the physical properties of Kepler-539\,c on a quite solid ground. While there are now several high-resolution spectrographs with RV performance better than CAFE, the observations of new transits of Kepler-539\,b are not straightforward with ground-based facilities. The 125.6-day orbital period and the long transit duration ($9.31$\,hr) put serious limitations for the successful of a such long-term monitoring and request a large observational effort with various telescopes located at different Earth's longitudes. Furthermore, the use of different telescopes could compromise the precision of the measurements of $T_0$ at the end. On the other hand, the observations of new mid-transit times of Kepler-539\,b can be easily performed by the incoming space telescope CHEOPS. This facility is therefore highly recommended for a new detailed follow-up study of the Kepler-539 planetary system.

\begin{acknowledgements}
This paper is based on observations collected with the 2.2\,m Telescope at the Centro Astron\'{o}mico Hispano Alem\'{a}n (CAHA) in Calar Alto (Spain) and the publicly available data obtained with the NASA space satellite \emph{Kepler}. Operations at the Calar Alto telescopes are jointly performed by the Max-Planck-Institut f\"{u}r Astronomie (MPIA) and the Instituto de Astrof\'{i}sica de Andaluc\'{i}a (CSIC). This research has been partially funded by Spanish grant AYA2012-38897-C02-01. J.L.-B. thanks the CSIC JAE-predoc program for Ph.D. fellowship. R.B.\ is supported by CONICYT-PCHA/Doctorado Nacional. R.B.\ acknowledges additional support from project IC120009 ``Millenium Institute of Astrophysics (MAS)'' of the Millennium Science Initiative, Chilean Ministry of Economy. We wish to thank Ennio Poretti for very helpful comments. We acknowledge the use of the following internet-based resources: the ESO Digitized Sky Survey; the TEPCat catalogue; the SIMBAD data base operated at CDS, Strasbourg, France; and the arXiv scientific paper preprint service operated by Cornell University.
\end{acknowledgements}

\bibliographystyle{aa}

%

\appendix
\section{Starspot parameters}
\label{appendix_A}
In this appendix we report a Table containing the starspot parameters that were determined from the {\sc prism}+{\sc gemc} fitting of the transit light curves at cycles $-5$ and $2$, as defined in Sect.\,\ref{sec_2.1}.

\begin{table*}
\centering 
\label{Table:starspotparameters}
\caption{Starspot parameters derived from the {\sc prism}+{\sc
gemc} fitting of the transit light curves at cycles $-5$ and $2$.
\newline{$^{a}$The longitude of the centre of the spot is
defined to be $0^{\circ}$ at the centre of the stellar disc and
can vary from $-90^{\circ}$ to $90^{\circ}$. $^{b}$The
co-latitude of the centre of the spot is defined to be $0^{\circ}$
at the north pole and $180^{\circ}$ at the south pole.
$^{c}$Angular radius of the starspot (note that an angular
radius of $90^{\circ}$ covers half of stellar surface).
$^{d}$Spot contrast (note that 1.0 equals the brightness of the
surrounding photosphere).}}%
\begin{tabular}{ccccc}
\hline
Cycle & ~~$\theta (^{\circ})\,^{a}$~~~ & $\phi(^{\circ})\,^{b}$ & $r_{\rm spot}(^{\circ})\,^{c}$ & $\rho_{\rm spot}\,^{d}$ \\
\hline
$-5$~~ & $-26.18 \pm ~\,3.73$~~ & $28.31 \pm 10.14 $ & $36.43 \pm 6.80$ & $0.69 \pm 0.11$ \\
$2$ & $18.88 \pm 11.79$ & $34.27 \pm ~\,8.21 $ & $18.60 \pm 6.46$ & $0.70 \pm 0.28$ \\
\hline
\end{tabular}
\end{table*}

\clearpage

\section{TRADES simulations}
\label{appendix_B}
In this appendix we report plots based on the results of the TRADES simulations for modelling the TTV detected in the transit timings of KOI-327\,b. In particular, the figures show RV plots (bottom panels) and O--C diagrams from linear ephemeris for planet Kepler-539\,b (top panels). In both cases, the observed data are compared with those obtained from the simulations (see Sect.\,\ref{sec_4}) The figures are ordered as in Table\,\ref{tab:trades}. Figures 1, 3 and 4 were obtained by using the PIKAIA algorithm \citep{charbonneau:1995}, while figures 2 and 5 with the Particle Swarm Optimization (PSO) algorithm \citep{tada:2007}.

\begin{figure}[th]
\resizebox{\hsize}{!}{\includegraphics[angle=0]{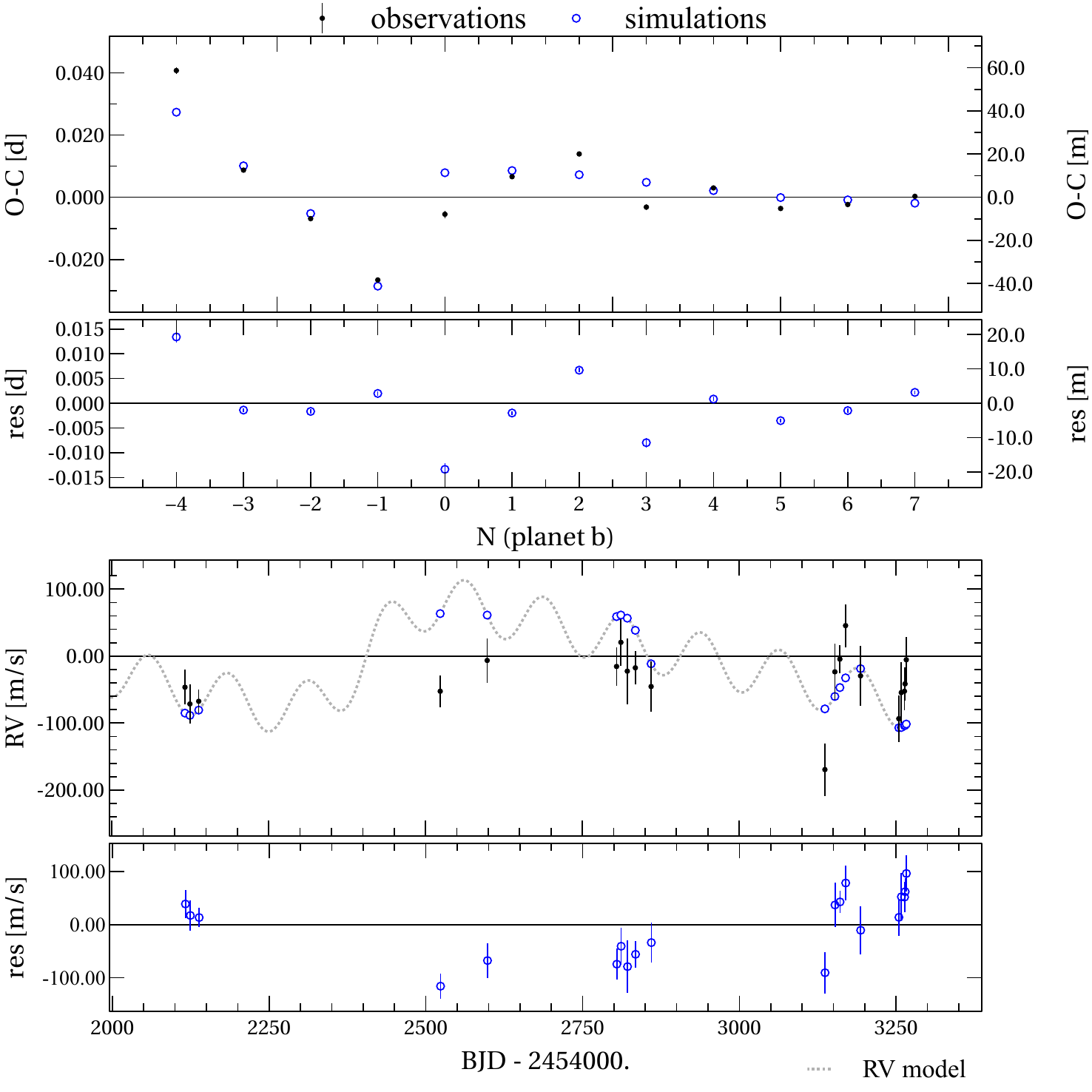}}
\caption{O--C diagram (with residuals) from linear ephemeris for
planet Kepler-539\,b (top panel); observations plotted as black
points (with error bars), simulations plotted as open blue
circles. Bottom panel shows the RV observations as solid black
circles, simulations at the same BJD(UTC) as open blue circles,
and the dotted gray line is the RV model for the whole simulation.
This the best solution found by {\sc trades}.}
\label{fig:simulationPIK1}
\end{figure}

\begin{figure}[th]
\resizebox{\hsize}{!}{\includegraphics[angle=0]{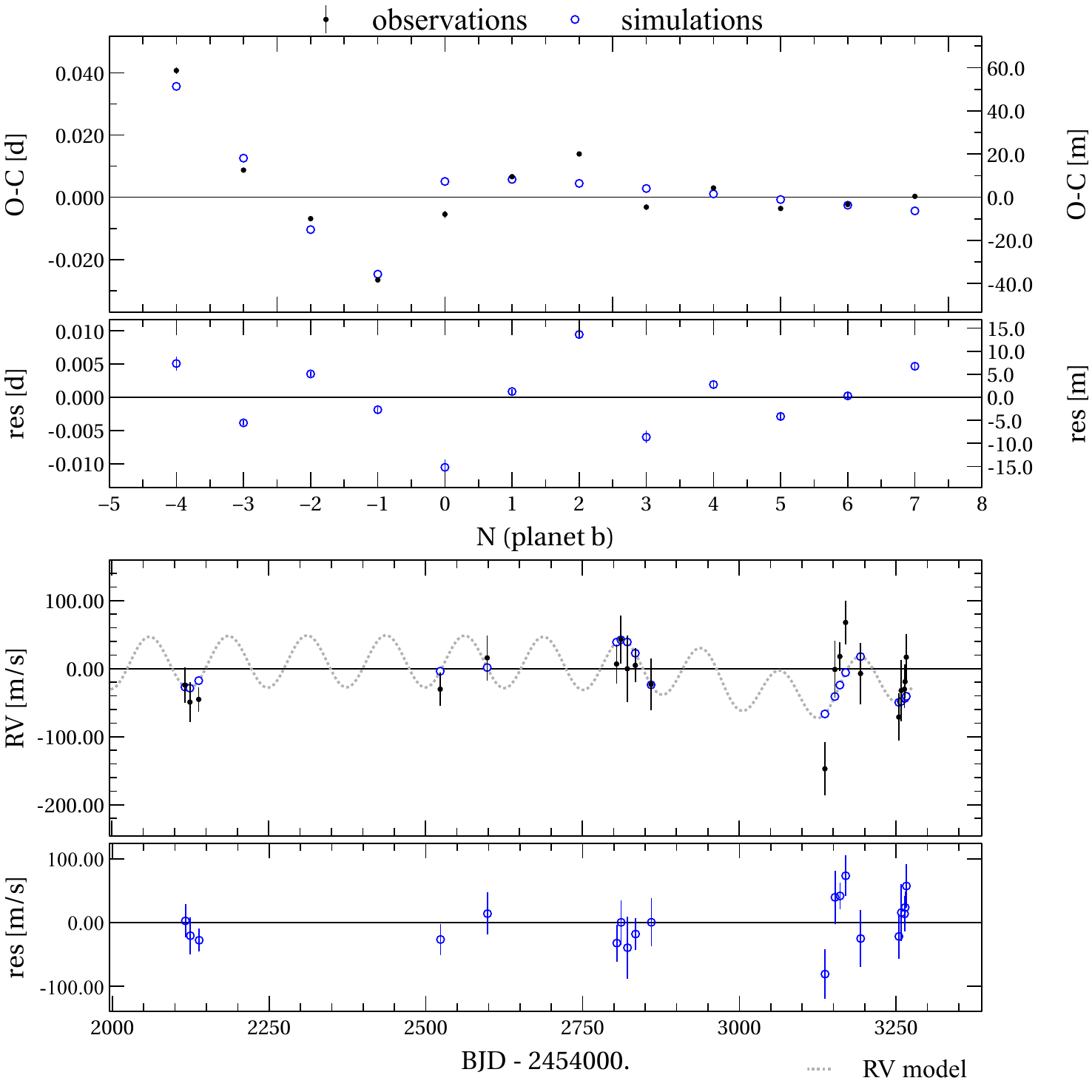}}
\caption{Same as for Fig\,\ref{fig:simulationPIK1}, but showing the 2$^{\rm nd}$ best solution found by {\sc trades}.}
\label{fig:simulationPSO2}
\end{figure}

\begin{figure}[th]
\resizebox{\hsize}{!}{\includegraphics[angle=0]{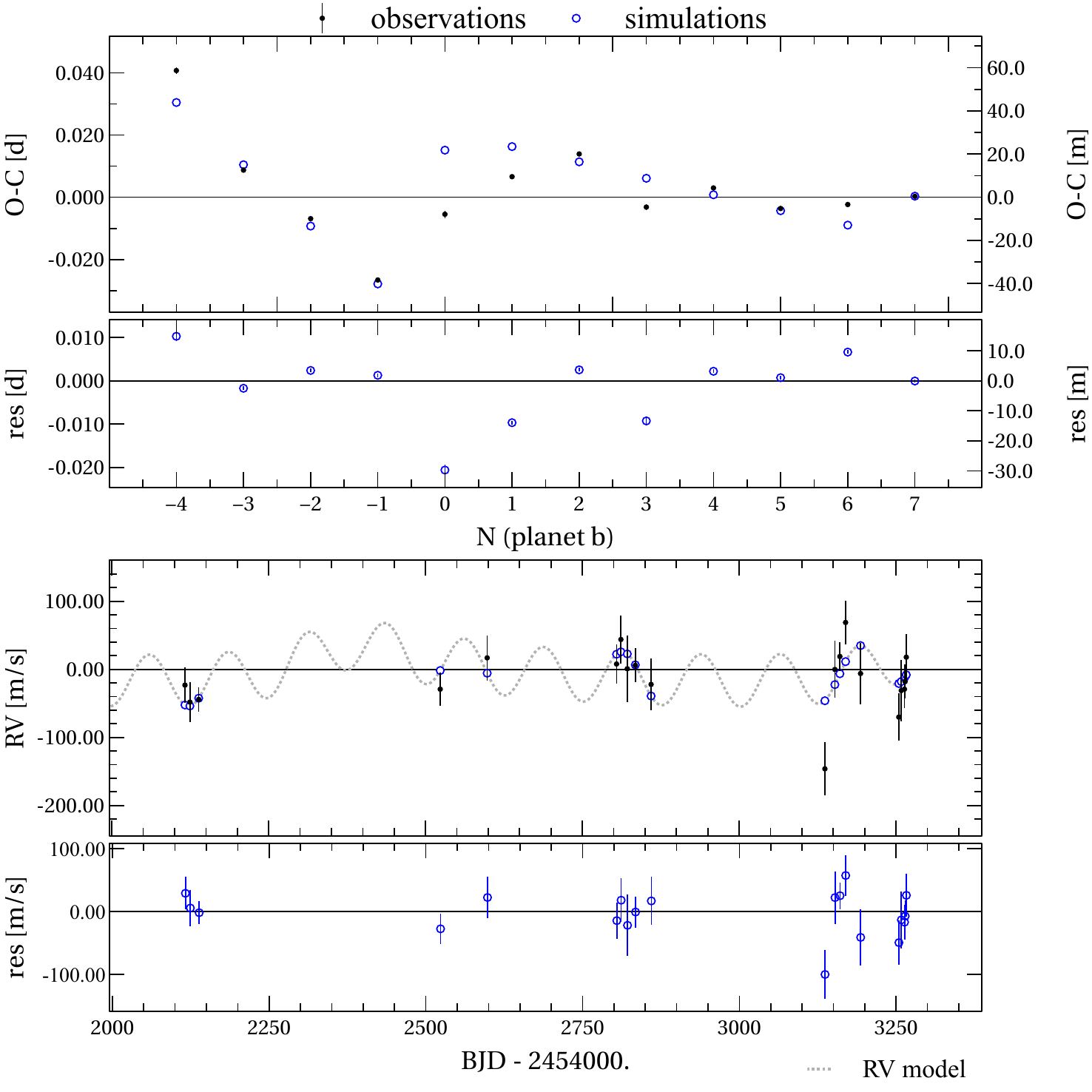}}
\caption{Same as for Fig\,\ref{fig:simulationPIK1}, but showing the 3$^{\rm rd}$ best solution found by {\sc trades}.}
\label{fig:simulationPIK2}
\end{figure}

\begin{figure}[th]
\resizebox{\hsize}{!}{\includegraphics[angle=0]{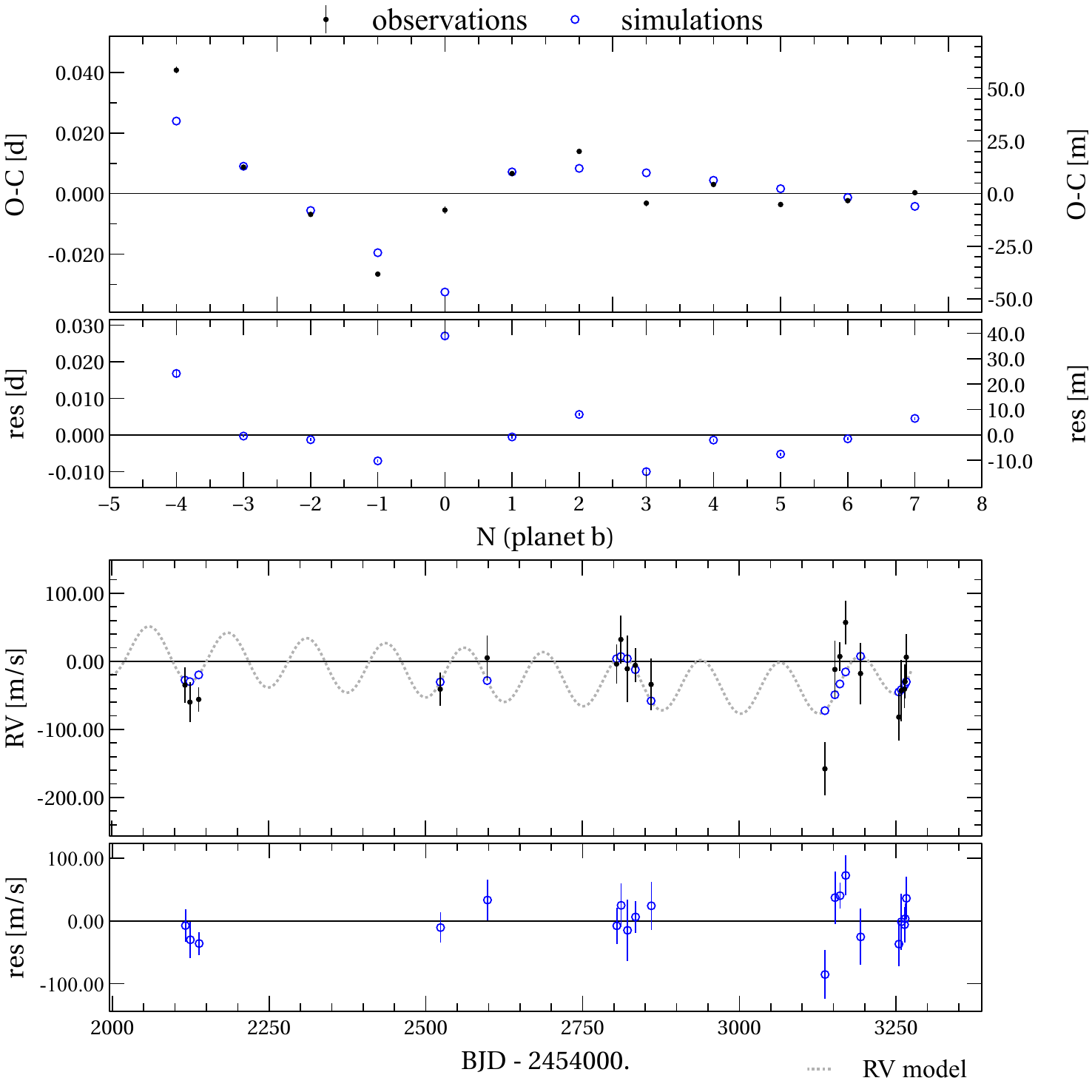}}
\caption{Same as for Fig\,\ref{fig:simulationPIK1}, but showing the 4$^{\rm th}$ best solution found by {\sc trades}.}
\label{fig:simulationPIK3}
\end{figure}

\begin{figure}[th]
\resizebox{\hsize}{!}{\includegraphics[angle=0]{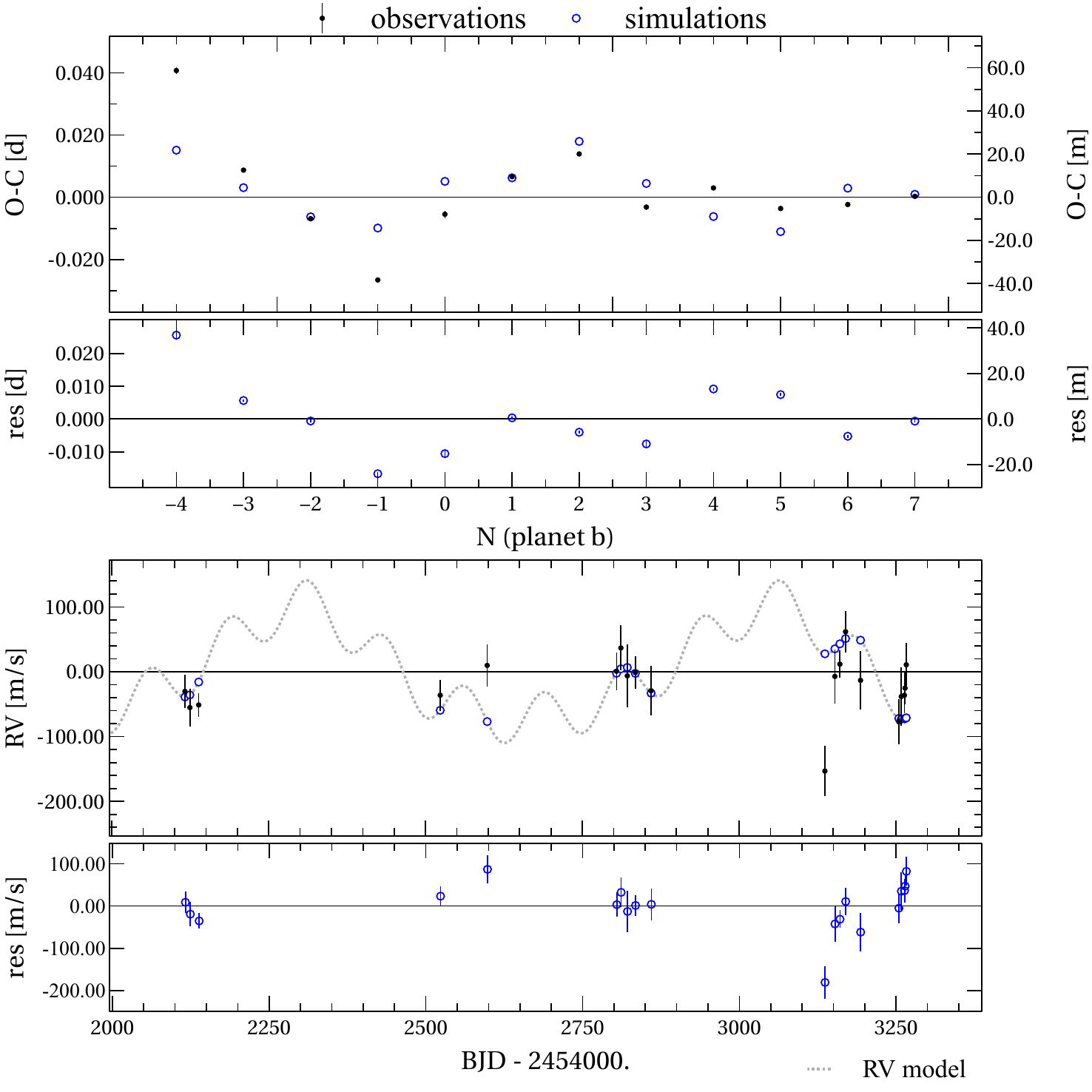}}
\caption{Same as for Fig\,\ref{fig:simulationPIK1}, but showing the 5$^{\rm th}$ best solution found by {\sc trades}.}
\label{fig:simulationPSO3}
\end{figure}

\end{document}